\documentstyle[aps,preprint,tighten]{revtex}
\begin{document}
\tolerance 10000
\title{
Supersymmetric electrodynamics \\
of charged and neutral fermions \\
in the extended Wheeler-Feynman approach}
\author{V.V.Tuga\u{\i}\footnote{Electronic Address: tugai@sptca.kharkov.ua}
}
\address{Scientific Center of Physical Technologies,
310145 Kharkov, Ukraine}
\author{A.A.Zheltukhin
\footnote{Electronic Address: kfti@rocket.kharkov.ua}
}
\address{Kharkov Institute of Physics and Technology,
310108 Kharkov, Ukraine}
\date{\today}
\maketitle

\begin{abstract}
A supersymmetric formulation of the classical action of
interacting charged and neutral fermions with arbitrary
anomalous magnetic moment is considered. 
This formulation generalizes the known action for scalar charged
particles investigated in papers by Fokker, Schwarzschild,
Tetrode, Wheeler and Feynman. 
The superfield formulation of the
electrodynamics of the Maxwell supermultiplet, constructed from
the world coordinates of charged or neutral fermions is carried
out basing on the proposed action.
\end{abstract}

\maketitle

The action at a distance approach developed by
Fokker-Schwarzschild-Tetrode-Wheeler-Feynman (FSTWF) revealed a deep
connection between classical electromagnetic field and world-line coordinates
of relativistic charged particles \cite{FW1}. The fundamental character of
the connections between fields and world coordinates was realized in a new
light in string theory where the original classical string lagrangian was
formulated in terms of string world-sheet coordinates \cite{SUPERSTRING}.
The unification of the FSTWF approach with string theory performed by Ramond
and Kalb resulted in the discovery of antisymmetric gauge field \cite{Kalb}.
Today this antisymmetric field plays an important role in supergravity,
superstring  and   modern topological theories.
Moreover, the cosmological studies of Hoyle and Narlikar \cite{Hoyle}
shed a new light on significance of the FSTWF approach
for the solving the old problem of the relation between microworld
physics and the features of the Universe as a whole.

As is known, the construction of successive field theory of string
interaction involves severe difficulties. We hope that FSTWF approach
may be useful for the construction of the theory of (super)string
interaction. In this connection the question about the possibility
of the unification of the AAD principle together with the principle
of supersymmetry seems the most
important question, since supersymmetry is the natural basis for
modern string theory construction.
By now, we solved this problem in a case of superparticles. The obtained
supersymmetrical generalization allows to elaborate the need acting
strategy and can form the basis for construction of (super)string
interaction.

The FSTWF approach is based on the action of the two
interacted charged particles in terms of its world-line coordinates
$x^{\mu }(t)$ and $y^{\mu }(\tau )$
\begin{equation}
S_{FSTWF} = e_1 e_2 \int dt \int d\tau \,
\dot x^\mu \dot y_{\mu} \delta (s^2_0).
\label{fokkerTwo}
\end{equation}
Here $s^{\mu}_0$ is the relativistic interval
and $\delta $ is the Dirac-function. The variation of this action together
with the ordinary free action gives the Lorentz equation of motion.
The strength of classical electromagnetic field
\begin {eqnarray}
a^{\mu} (x)=e \int d{\tau} {\dot y}^{\mu}(\tau ) \delta (s^2 _0)
\label{a_clas}
\end{eqnarray}
appears in the right hand side of this equation.
This field satisfies the Maxwell equation with the current,
which is given in a standard way
\begin {eqnarray}
&&{\partial}^{\mu} f_{\mu \nu}(x)=-4{\pi}j_{\nu}(x),\ \ \ \
j^{\mu}(x)=e\int d\tau {\dot y}^{\mu}{\delta}^{(4)}(s_0)
\label{maxv_cl}
\end {eqnarray}
and the Lorentz gauge condition
\begin{eqnarray}
&&{\partial}^{\mu } a_{\mu}(x)=0.
\label{clLor}
\end{eqnarray}

In order to supersymmetrize the electromagnetic potential (\ref{a_clas}) and
the current we extend the original Minkowski space to
superspace.
$z^M=(x^{\mu }(t),{\theta }^{\alpha }(t),{\bar \theta }_{\dot \alpha }(t))$
and ${\zeta}^M=(y^{\mu }(\tau ),{\xi }^{\alpha }(\tau ),
{\bar \xi }_ {\dot \alpha }(\tau ))$ are
the superspace coordinates of particles, where additional
Grassmann coordinates $\theta $ and $\xi$
are the Weyl spinors. The supersymmetry transformations mix
the ordinary vector and spinor coordinates of particles \cite{WESS}
\begin{eqnarray}
{\delta}x^{\mu }=i{\theta}{\sigma}^{\mu}{\bar{\epsilon}}-
i{\epsilon}{\sigma}^{\mu}{\bar{\theta}}, \quad
{\delta}{\theta}^{\alpha}={\epsilon}^{\alpha} , \
{\delta}{\bar{\theta}}_{\dot \alpha}={\bar{\epsilon}}_{\dot \alpha} \nonumber
\\
{\delta}y^{\mu }=i{\xi }{\sigma}^{\mu}{\bar{\epsilon}}-
i{\epsilon}{\sigma}^{\mu}{\bar{\xi }}, \quad
{\delta}{\xi }^{\alpha}={\epsilon}^{\alpha} , \
{\delta}{\bar{\xi }}_{\dot \alpha}={\bar{\epsilon}}_{\dot \alpha}.
\label{SUSYtrans}
\end{eqnarray}
The simplest generalizations of the interval $s_0^{\mu} $ and the velocity
${\dot y}^{\mu}(\tau )$ invariant under the global supersymmetry
transformations (\ref{SUSYtrans}) are
\begin{eqnarray}
&& s^{\mu}=x^{\mu}-y^{\mu}-i({\theta}{\sigma}^{\mu}{\bar{\xi }}
-{\xi } {\sigma}^{\mu}{\bar{\theta}}), \nonumber\\
&&{\omega}^{\mu}_{\tau}={\dot {y}}^{\mu}-
i( {\dot \xi }{\sigma}^{\mu}{\bar{\xi }}-
{\xi }{\sigma}^{\mu} {\dot {\bar{\xi }}}).
\label{subst2}
\end{eqnarray}

However, as will be shown further, it is more convenient to work
with the basis of chiral coordinates
$(x^{\mu}_L, \theta ^{\alpha })$, $(x^{\mu}_R, \bar\theta _{\dot\alpha })$
\cite{WESS}.
\begin{eqnarray}
x^{\mu}_L {\equiv} x^{\mu}+ i{\theta}{\sigma}^{\mu}{\bar \theta},\quad
y^{\mu}_L {\equiv} y^{\mu}+ i{\xi }{\sigma}^{\mu}{\bar \xi },
\nonumber \\
x^{\mu}_R {\equiv} x^{\mu}- i{\theta}{\sigma}^{\mu}{\bar \theta},\quad
y^{\mu}_R {\equiv} y^{\mu}- i{\xi }{\sigma}^{\mu}{\bar \xi }
\label{xyLR}
\end{eqnarray}
Then it is not so difficult to construct the intervals
\begin{eqnarray}
&& s_L ^{\mu}=x_L ^{\mu}-y_R ^{\mu}-2i{\theta}{\sigma}^{\mu}{\bar{\xi }}
=s^{\mu}+i\Delta{\sigma }^{\mu }\bar{\Delta},\quad
{\Delta}^{\alpha } ={\theta}^{\alpha}-{\xi}^{\alpha },
\nonumber \\
&& s_R ^{\mu}=x_R ^{\mu}-y_L ^{\mu}+2i{\xi}{\sigma}^{\mu}{\bar{\theta}}
=s^{\mu}-i\Delta{\sigma }^{\mu }\bar{\Delta},\quad
{\bar{\Delta}}^{\dot{\alpha}}
={\bar\theta}^{\dot\alpha}-{\bar\xi}^{\dot\alpha },\nonumber\\
\label{substCIR}
\end{eqnarray}
which are invariant under the SUSY transformations (\ref{SUSYtrans})
and satisfy the chirality conditions
\begin{eqnarray}
&&D_{\alpha}s_R^{\mu} = \left( {\partial \over \partial {\theta}^{\alpha} }
+i({\sigma }^{\nu }{\bar\theta})_{\alpha }\partial _{\nu } \right) s_R^{\mu }
=D_{\alpha}{\bar\Delta _{\dot\beta}}=0,\nonumber\\
&&{\bar D}_{\dot\alpha}s_L^{\mu} = \left(- {\partial \over
\partial {\bar\theta}^{\dot\alpha}}
-i({\theta}{\sigma }^{\nu })_{\dot\alpha} \partial _{\nu } \right) s_L^{\mu }
={\bar D}_{\dot\alpha}\Delta ^{\beta}=0.
\label{CondCIR}
\end{eqnarray}

The main objects of the supersymmetric electodynamics are
the superfield vector and spinor connections
$A^M = (A^{\mu }, A^{\alpha}, {\bar A}_{\dot\alpha})$.
These potentials play the same role as the
vector electromagnetic potential in the standard electrodynamics.
The corresponding strength components $F_{MN}$ are not
independent superfields due to the standard constraints \cite{WESS}
\begin{eqnarray}
F_{\alpha \beta }=F_{\dot{\alpha }\dot{\beta }}=F_{\alpha {\dot \beta}}=0.
\label{ConstrWess}
\end{eqnarray}
Then all nonzero components of the strength $F_{MN}$ can be expressed
through the chiral superfields $W_{\alpha }$ and its complex conjugated
ones
\begin{eqnarray}
&& W^{\alpha }\equiv
{i \over 4} F_{\mu {\dot \alpha }}{\tilde \sigma }
^{\mu {\dot \alpha}\alpha}, \quad
{\bar W}^{\dot \alpha }
\equiv
{i \over 4} {\tilde \sigma }^{\mu {\dot \alpha}\alpha} F_{\mu \alpha }.
\label{WfromF}
\end{eqnarray}

Standard constraints are automatically satisfied if the vector superfield
$A_{\mu }$ has the form
\begin{eqnarray}
A_{\mu }= -{i\over 4}
{\tilde\sigma}_{\mu}^{\dot\alpha {\alpha }} \left(
D_{\alpha }{\bar A}_{\dot \alpha  }+{\bar D}_{\dot \alpha  } A_{\alpha }
\right)
\label{VectFrSpinPot}
\end{eqnarray}
and the spinor superfields $A^{\alpha}$ and ${\bar A}_{\dot\alpha}$ are
chiral
\begin{eqnarray}
D_{\alpha}A_{\beta }=0, \quad {\bar D}_{\dot\alpha}{\bar A}_{\dot \beta}=0.
\label{CIRpot}
\end{eqnarray}

The above expressions together with the conjugation condition
for the spinor superfields allow to reduce the problem of construction of the
superpotentials $A^M$ to the problem of obtaining the spinor
superfield $A^{\alpha }$.

Let us seek for an integral representation for spinor potential
in the form
\begin{eqnarray}
A_{\alpha}(x,\theta ,{\bar\theta}) =\int d\tau {\cal K}_{\alpha}
({\omega }_{\tau},{\dot\xi},{\dot{\bar\xi}},\bar\Delta )
\delta(s^{2}_R ),
\label{PotspinFrKer}
\end{eqnarray}
where the kernel ${\cal K}_{\alpha }$ is the chiral supersymmetric
invariant operator and has a dimensionality ${L}^{1\over 2}$ in system
$h=c=1$.

Then spinor strength $W$ can be rewritten by the following way
\cite{WESS}
\begin{eqnarray}
{W}^{\alpha }&=& {1\over 2} \int d\tau \biggl[
{1\over 4}{\bar D}{\bar D}({\cal K}^{\alpha })-i(\Delta \sigma ^{\rho }
{\bar D})({\cal K}^{\alpha})\partial _{\rho }-{\cal K}^{\alpha }
(\Delta \Delta ) \Box \biggr] \delta (s_R^2 )
\nonumber\\
&&
-{i\over 2} \int d\tau
{\cal K}^* _{\dot \alpha}\partial ^{\dot\alpha \alpha }\delta (s_L^2 ).
\label{WfromK}
\end{eqnarray}

We use the superfields $W$ and  ${\bar W}$ in order to construct
the representation for supersymmetric current
\begin{eqnarray}
-4\pi {\cal J} =D^{\alpha }W_{\alpha }+
{\bar D}_{\dot\alpha }{\bar W}^{\dot \alpha }.
\label{TOKtoPOT}
\end{eqnarray}
Further we assume that the following
condition for kernel $\cal K$
\begin{eqnarray}
({\bar D}\sigma ^{\mu }{\cal K}) \sim {d\over d\tau } (s_R^{\mu })
\label{TOKcondKER}
\end{eqnarray}
is satisfied.
This condition plays the same role, as the Lorentz gauge condition in the
original FSTWF theory --- it  permits to write the above
expression in the form of wave equation
\begin{eqnarray}
\Box \Phi (x,\theta,{\bar \theta})
&=&-4{\pi} {\cal J}(x,\theta,{\bar \theta}), \ \hbox{where}
\cr
\Phi  &=& -2
\int d\tau\left(    {\Delta}^{\alpha }{\cal K}_{\alpha }\delta (s_R^2)+
{\cal K}^*_{\dot\alpha } {\bar\Delta}^{\dot\alpha }\delta (s_L^2)\right) .
\label{PredpotFrKer}
\end{eqnarray}
As one can see, the fundamental property of the interval
\begin{eqnarray}
\Box \delta(s^2)=-4{\pi}{\delta}^{(4)}
(s^{\nu })
\label{DIRACclas}
\end{eqnarray}
gives the possibility to write an integral representation for the
supersymmetric current
\begin{eqnarray}
{\cal J} &=& -4
\int d\tau   {\Delta}^{\alpha }{\cal K}_{\alpha }\delta ^{(4)}(s_R)+
{\cal K}^*_{\dot\alpha }{\bar\Delta}^{\dot\alpha }\delta^{(4)}(s_L).
\label{FIELDeqSUSY}
\end{eqnarray}
This wave equation is a superfield generalization of the Maxwell
equations in the Lorentz gauge. The superfield
$\Phi $ has the physical meaning of the prepotential
$V$ evaluated in the superfield gauge
\begin{eqnarray}
\left\{ DD,{\bar D}{\bar D} \right\} V =0 \ \Rightarrow
V(x,\theta,{\bar \theta})= {1\over 4}\Phi (x,\theta,{\bar \theta}).
\label{calibrSUSYtwo}
\end{eqnarray}
The superfield gauge condition (\ref{calibrSUSYtwo})
is split into the following component gauge conditions
\begin{eqnarray}
&&{\partial }^{{\dot\alpha }\alpha } {\chi}_{\alpha }(x)=
i{\bar \lambda }^{\dot \alpha }(x),\
{\partial }^{{\dot\alpha }\alpha } {\bar \chi}_{\dot \alpha }(x)=
-i{\lambda }^{\alpha }(x),
\nonumber\\
&&\Box C(x)=-D(x),\
M(x)=N(x)=0,
\nonumber \\
&&{\partial }_{\mu }v^{\mu }(x)=0.
\label{calibrCONDcomp}
\end{eqnarray}
Here $v^{\mu }(x)$, ${\lambda}^{\alpha }(x) $ and $D(x)$ ---
are the Maxvell supermultiplet fields, $\chi$, $M$, $N$, $C$ ---
are the component fields, which vanish in Wess-Zumino gauge.
Above gauge conditions (\ref{calibrCONDcomp})
permit to represent  prepotential $V(x,\theta,{\bar \theta})$ as
the following component decomposition
\begin{eqnarray}
V(x,\theta, \bar\theta)&=&
-\Box ^{-1}D- {\theta }^{\alpha } (\partial ^{-1})_{\alpha\dot\beta}
\lambda ^{\dot\beta} -i{\bar \theta }_{\dot \alpha }
(\partial ^{-1})^{\dot\alpha \beta}
\lambda _{\beta} -
(\theta {\sigma }_{\rho }{\bar \theta}){v}^{\rho}
\nonumber\\
&&+{i\over 2}{\theta}{\theta } {\bar \theta }_{\dot \alpha }
{\bar\lambda }^{\dot\alpha }
-{i\over 2}{\bar \theta }{\bar\theta } {\theta }^{\alpha } {\lambda }_{\alpha}
+{1\over 4}{\theta}{\theta } {\bar \theta }{\bar \theta } D
\label{MAXWpreptwo}
\end{eqnarray}
with the vector $v^{\mu }(x)$, two spinors ${\lambda}^{\alpha }(x) $,
${\bar\lambda}_{\dot\alpha }(x) $ and an auxiliary scalar field $D(x)$
forming the Maxwell multiplet.

The components of supercurrent can be defined as
\begin{eqnarray}
{\cal J} &=& -4 {j}^{(0)}+4{\theta }^{\alpha }
{j}_{\alpha  } ^{(1)}-4{\bar \theta }_{\dot \alpha }{\bar {j}}^
{{\dot \alpha}(1)}-4(\theta {\sigma }_{\rho }{\bar \theta}){ j}^{(2)\rho }
\nonumber\\
&& -2i{\theta}{\theta } {\bar \theta }_{\dot \alpha }{\partial }^
{{\dot \alpha }\alpha }{j}_{\alpha  } ^{(1)}+2i{\bar \theta }{\bar\theta }
{\theta }^{\alpha }{\partial }_{\alpha {\dot \alpha }}
{\bar {j}}^{(1) \dot \alpha } +\theta \theta {\bar \theta }{\bar \theta }
\Box {j}^{(0)}.
\nonumber \\
\label{TOKtocomp}
\end{eqnarray}
After substitution of both of these decompositions into the superfield
wave equation one can obtain
that the component fields of $\Phi $ satisfy the Maxwell
and Dirac equations with currents
\begin{eqnarray}
&&\Box v^{\mu}(x)=-4\pi{j}^{(2)\mu }(x),\
{\partial }_{\alpha {\dot \alpha }}{\bar \lambda }^{\dot \alpha }(x)=
-4\pi{j}_{\alpha }^{(1)}(x).
\label{compFIELDeq}
\end{eqnarray}

Now we consider the representations for kernel in terms
of the intervals and velocities. These quantities are chiral and
supersymmetry invariant ones. Thus the constructed kernel
\begin{eqnarray}
{\cal K}^{(e)}_{\alpha}=e {\sigma }^{\mu }_{\alpha \dot\alpha }
\bigl(
{\omega }_{\tau \mu}+2i
({\bar\Delta }{\tilde\sigma }_{\mu }\dot\xi) \bigr)
{\bar\Delta }^{\dot\alpha },
\label{POTspinTWO}
\end{eqnarray}
is chiral and supersymmetry invariant operator too.
It is not so difficult to verify that this kernel satisfies
the above introduced condition for kernel and has the appropriate dimensionality.
Note, that the corresponding vector superpotential
satisfies the superfield Lorentz gauge condition
$$
{\partial }_{\mu }A^{\mu }=0.
$$
If we substitute this representation of kernel ${\cal K}$ into
expressions for superfield potentials, strengths and current
they can be rewritten in terms of superspace coordinates \cite{ZT3}
\begin{eqnarray}
A_{\alpha}&=&e \int d\tau \bigl(
{\omega }_{\tau \mu}{\sigma }^{\mu }_{\alpha \dot\alpha }{\bar\Delta }^
{\dot\alpha } +2i{\dot\xi}_{\alpha }{\bar\Delta}{\bar\Delta} \bigr)
\delta(s^{2}_R ), \nonumber\\
{\bar A}_{\dot\alpha} &=& -e \int d\tau \bigl(
{\omega }_{\tau \mu}{\Delta }^{\alpha }{\sigma }^{\mu }_{\alpha \dot\alpha }
-2i{\dot{\bar\xi}}_{\dot\alpha }{\Delta}{\Delta} \bigr) \delta(s^{2}_L ),\cr
A_{\mu }&=&
-i e \int d\tau \biggl[ {\omega }_{\tau \mu} -{\varepsilon}_{\mu \nu \rho
\lambda } {\omega }_{\tau }^{\nu}(\Delta{\sigma}^{\rho}{\bar\Delta})
{\partial }^{\lambda }
\nonumber \\
&&
+i\Bigl( ({\Delta}{\sigma }_{\mu }{\dot{\bar\xi}})
-({\dot\xi}{\sigma }_{\mu }{\bar\Delta}) \Bigr) +
{1\over 4}{\Delta}{\Delta }{\bar\Delta }{\bar\Delta }
{\omega }_{\tau}^{\nu}\Bigl( {\partial}_{\mu}{\partial}_{\nu}
-{\eta}_{\mu\nu}\Box \Bigr)
\nonumber \\
&&+\left( {\Delta}{\Delta} ({\dot{\bar\xi}}{\tilde\sigma }_{\mu \rho }
{\bar\Delta }) +({\dot\xi}{\sigma }_{\mu \rho }
{\Delta}){\bar\Delta}{\bar\Delta} \right)
{\partial}^{\rho } \biggr] \delta (s^2),
\label{vecttospinPOT}
\end{eqnarray}
\begin{eqnarray}
{W}^{\alpha }
&=& -i e \int d\tau \biggl[
{\dot \xi}^{\alpha } +
i{\dot \xi}^{\alpha } \Delta {\sigma}^{\mu } {\bar\Delta} {\partial }_{\mu }
+{1\over 4}{\dot\xi}^{\alpha}\Delta\Delta{\bar\Delta}{\bar\Delta}\Box
\nonumber\\
&&+{\omega}_{\tau \mu}
\Bigl(
2(\Delta {\sigma }^{\mu \nu })^{\alpha }{\partial }_{\nu }
-{i\over 2} {\Delta}{\Delta} ({\bar\Delta}{\tilde\sigma }_{\nu })^{\alpha }
\bigl( {\partial }^{\mu } {\partial }^{\nu },
-{\eta}^{\mu \nu }\Box \bigr)\Bigr)
\nonumber\\
&&-{i} {\Delta}{\Delta} ({\dot{\bar\xi}}{\tilde\sigma }_{\mu })^{\alpha }
{\partial }^{\mu }
\biggr] \delta (s^2),\cr
\Phi  &=& -2e \int d\tau \Bigl(
{\omega }_{\tau }^{\mu}\bigl( {\Delta}{\sigma }_{\mu }{\bar\Delta }\bigr)
+i\bigl( {\dot\xi}\Delta\bigr) {\bar\Delta}{\bar\Delta}
-i{\Delta}{\Delta} \bigl( {\dot{\bar\xi}}{\bar \Delta}\bigr) \Bigr)
\delta (s^2).
\label{WfromY}
\end{eqnarray}
The integral representations of the component fields in terms of the
superspace coordinates can be written as zero-terms of the corresponding
superfields
\begin{eqnarray}
v_{\mu }(x)&=&iA_{\mu }\Bigl|_{{\theta}=0}
=e \int d\tau \biggl[ {\dot y}_{\mu} -{\varepsilon}_{\mu \nu \rho \lambda }
{\dot y}^{\nu}(\xi{\sigma}^{\rho}{\bar\xi}){\partial }^{\lambda }
\nonumber \\
&&+\left( {\xi}{\xi} ({\dot{\bar\xi}}{\tilde\sigma }_{\mu \rho }{\bar\xi })+
({\dot\xi}{\sigma }_{\mu \rho }{\xi}){\bar\xi}{\bar \xi} \right)
{\partial}^{\rho }
+{1\over 4}{\xi}{\xi} {\bar\xi}{\bar \xi} {\dot y}^{\nu }
\left( {\partial }_{\mu}{\partial }_{\nu}-{\eta}_{\mu\nu}\Box \right)
\biggr] \delta (s_0^2),
\cr
{\lambda }^{\alpha }(x)
&=&iW_{\alpha  }\Bigl|_{{\theta}=0}
= e \int d\tau \biggl[ {\dot \xi}^{\alpha } -
i{\dot\xi}{\xi}({\bar\xi}{\tilde \sigma }_{\mu })^{\alpha }{\partial }^{\mu }
+{i\over 2} {\xi}{\xi} ({\dot{\bar\xi}}{\tilde\sigma }_{\mu })^{\alpha }
{\partial }^{\mu }
\nonumber\\
&&-{1\over 2}{\dot \xi}^{\alpha }{\xi}{\xi}{\bar \xi}{\bar \xi}\Box
+{\dot y}_{\mu}
\biggl(
-2(\xi {\sigma }^{\mu \nu })^{\alpha }{\partial }_{\nu }
+{i\over 2} {\xi}{\xi} ({\bar\xi}{\tilde\sigma }_{\nu })^{\alpha }
\bigl( {\partial }^{\mu } {\partial }^{\nu }- {\eta}^{\mu \nu }\Box \bigr)
\biggr)
\biggr] \delta (s_0^2),
\cr
D(x)
&=&-{1\over 4}\Box \Phi \Bigl|_{{\theta}=0}
= e \int d\tau
\biggl[
{\dot y}_{\mu}(\xi{\sigma }^{\mu }{\bar\xi})
-i\left( {\xi}{\xi} ({\dot{\bar\xi}}{\bar\xi })
-({\dot\xi}{\xi}){\bar\xi}{\bar\xi} \right)
\biggr] \Box \delta (s_0^2).
\label{compVECTtwo}
\end{eqnarray}
It is obvious, that the electromagnetic potential $v^{\mu }(x)$
is the desired supersymmetric generalization of the original FSTWF potential.

The differential operators of D'Alambert and Dirac applied to the integral
representations for the component fields allow to obtain the explicit form
for the components (\ref{TOKtocomp}) of the current multiplet
${\cal J}(x,\theta ,\bar\theta )$ 
\begin{eqnarray}
j^{(2)}_{\mu }&=&
e \int d\tau \biggl[ {\dot y}_{\mu} -{\varepsilon}_{\mu \nu \rho \lambda }
{\dot y}^{\nu}(\xi{\sigma}^{\rho}{\bar\xi}){\partial }^{\lambda }+
{1\over 4}{\xi}{\xi} {\bar\xi}{\bar \xi} {\dot y}^{\nu }
\bigl( {\partial }_{\mu }{\partial }_{\nu }
\nonumber\\
&&- {\eta}_{\mu \nu } \Box \bigr)
+\left( {\xi}{\xi} ({\dot{\bar\xi}}{\tilde\sigma }_{\mu \rho }{\bar\xi })
+({\dot\xi}{\sigma }_{\mu \rho }{\xi}){\bar\xi}{\bar \xi} \right)
{\partial}^{\rho } \biggr] \delta ^{(4)}(s_0), \cr
j^{(1)}_{\alpha } &=&e \int d\tau \biggl[ {\dot y}_{\mu}
\bigl[ ({\sigma }^{\mu }{\bar\xi})_{\alpha }
-{i\over 2} {\xi}_{\alpha }{\bar \xi}{\bar \xi} {\partial }^{\mu }
-i({\sigma }^{\mu \rho }\xi)_{\alpha }{\bar \xi}{\bar \xi}{\partial }_{\rho }
\bigr] \nonumber\\
&&+i {\dot\xi}_{\alpha }{\bar\xi}{\bar \xi}
-{1\over 2}({\sigma}^{\rho}{\dot{\bar\xi}})_{\alpha}{\xi}{\xi}
{\bar \xi}{\bar \xi}{\partial}_{\rho } \biggr] \delta ^{(4)}(s_0).
\label{TOCcompVECTtwo}
\end{eqnarray}

Now we choose another representation of kernel $\cal K$
\begin{eqnarray}
{\cal K}^{(\mu )}_{\alpha }=\mu \left( {\dot \xi}_{\alpha} -
i {\dot{\bar {\xi}^{\dot \alpha}}}{\bar\Delta}{\bar\Delta}
{\partial }_{\alpha {\dot\alpha }}\right) ,
\label{Kmu}
\end{eqnarray}
where the constant $\mu $ has the dimensionality of the length $L$
and has a physical meaning of the anomalous magnetic moment (AMM) of a
neutral source particle \cite{ZT1}.
This new representation satisfies the same conditions for the desired
kernels as the above mentioned one.
The obtained fields are generated by superparticles with nonzero AMM.

The coordinate integral representations of the main superfields
and component fields can be written in the same manner as for the
charged superparticle-source.
\begin{eqnarray}
A^{\alpha} &=&\mu  \int d\tau
\left[ {\dot \xi}^{\alpha} -i {\dot{\bar {\xi}^{\dot \alpha}}}
{\bar\triangle}{\bar\triangle}{\partial }_{\alpha {\dot\alpha }}
\right] \delta(s^{2}_R ),
\cr
A_{\nu } &=&\mu  \int d\tau
\Bigl\{
2\left[ {\dot \xi}{\sigma }_{\mu \nu }\triangle -
{\bar\triangle}{\tilde\sigma }_{\nu \mu} {\dot{\bar \xi}}\right] {\partial }^{\mu }+
\nonumber\\
&&+{i\over 2}\left[ {\triangle}{\triangle}{\dot\xi}{\sigma}^\rho {\bar\triangle}
+{\bar\triangle}{\bar\triangle}{\triangle}{\sigma }^{\rho } {\dot{\bar\xi}}
\right] ({\eta}_{\rho \nu }\Box-{\partial }_{\rho }{\partial }_{\nu })
\Bigr\} \delta(s^{2} ),
\cr
{W}^{\alpha } &=& -i \mu  \int d\tau \Bigl[
{\dot {\bar{\xi}_{\dot\alpha }}}{\partial }^{{\dot\alpha}\alpha}
\delta (s_L^2) -i
{\dot \xi}^{\alpha }{\Delta}{\Delta}\Box \delta (s_R^2) \Bigr] ,
\cr
\Phi  &=& -4\mu\int d\tau \left[
\dot\xi ^{\alpha }\Delta _{\alpha }\delta (s_R^2)+
{\dot{\bar\xi}}_{\dot\alpha}{\bar\triangle}^{\dot\alpha}
\delta(s_L^{2})\right],
\cr
v_{\nu }&=& \mu \int d\tau
\Bigl[
-2i\left( {\dot \xi}{\sigma }_{\mu \nu }\xi -
{\bar\xi}{\tilde\sigma }_{\nu \mu} {\dot{\bar \xi}}\right) {\partial}^{\mu}
\cr
&&+{1\over 2}\left( {\xi}{\xi}{\dot\xi}{\sigma}^\rho {\bar\xi}
+{\bar\xi}{\bar\xi}{\xi}{\sigma }^{\rho } {\dot{\bar\xi}}
\right) ({\eta}_{\rho \nu }\Box-{\partial }_{\rho }{\partial }_{\nu })
\Bigr] \delta(s_0^{2} ),
\label{compVECTone}
\\
{\lambda }^{\alpha } &=& \mu  \int d\tau \left[
-i {\dot \xi}^{\alpha }{\xi}{\xi}\Box
+{\dot {\bar{\xi}_{\dot\alpha }}}\left(
{\partial }^{{\dot\alpha}\alpha}+i
{\xi}^{\beta }{\bar\xi}^{\dot\beta}
{\partial }^{{\dot\alpha }\alpha }{\partial }_{{\dot\beta }\beta }+
{1\over 4}{\xi}{\xi}{\bar\xi}{\bar\xi}
{\partial }^{{\dot\alpha }\alpha }\Box \right) \right]\delta (s_0^2),
\label{compSPINone}
\\
D& =& -{1\over 2}\mu  \int d\tau
\left[ 2 ({\dot\xi}{\xi}+ {\dot{\bar\xi}}{\bar\xi}) +
i\left( {\xi}{\xi}{\dot\xi}{\sigma }^{\mu }{\bar\xi}-
{\bar\xi}{\bar\xi}{\xi}{\sigma }^{\mu }{\dot{\bar\xi}}
\right) {\partial }_{\mu }
\right] \Box \delta(s_0^{2}).
\label{compDone}
\\
j^{(2)}_{\nu }&=&
\mu \int d\tau \Bigl[
-2i\left( {\dot \xi}{\sigma }_{\mu \nu }\xi -
{\bar\xi}{\tilde\sigma }_{\nu \mu} {\dot{\bar \xi}}\right)
{\partial}^{\mu}
\cr
&&+{1\over 2}\left( {\xi}{\xi}{\dot\xi}{\sigma}^\rho {\bar\xi}
+{\bar\xi}{\bar\xi}{\xi}{\sigma }^{\rho } {\dot{\bar\xi}}
\right) ({\eta}_{\rho \nu }\Box-{\partial }_{\rho }{\partial }_{\nu })
\Bigr] \delta ^{(4)}(s_0),
\label{TOCcompVECTone}
\\
j^{(1)}_{\alpha }& =&
\mu  \int d\tau \left[ i {\dot{\bar\xi}}^{\dot\alpha }{\bar\xi}{\bar\xi}
{\partial}_{\alpha {\dot\alpha }}
-{\dot \xi}_{\alpha }\left( 1-i{\xi}{\partial}{\bar\xi}
+{1\over 4}{\xi}{\xi}{\bar\xi}{\bar\xi}\Box \right) \right] \delta ^{(4)}(s_0).
\label{TOCcompSPINone}
\end{eqnarray}

The next important step of our work is
the construction of the action functional for the interacting superparticles
with charge and AMM.
In the case of two interacting charged superparticles
the corresponding action functional
can be written as a sum of free and interaction part,
which are defined by the expressions
\begin{eqnarray}
S_0 &=&-{1\over 2}\int dt \Bigl( {\textstyle \omega _t^2\over
\textstyle g_t}
+g_t m_1^2 \Bigr) -{1\over 2}\int d\tau \Bigl(
{\textstyle \omega _{\tau}^2\over \textstyle g_{\tau}} +g_{\tau} m_2^2
\Bigr),
\nonumber\\
S_{int}^{(e)}&=& i\, e \int dt \Bigl( {{\omega}}^{\mu}_t A_{\mu}
+{\dot {\theta}}_t^{\alpha} A_{\alpha}
+{\dot {\bar{\theta}}}_{t\dot \alpha} {\bar A}^{\dot {\alpha}}  \Bigr)
=i e \int \omega ^{M} (d)  A_M .
\label{ActionSUSYIntAndFree}
\end{eqnarray}
Here, instead of the gauge superfield $A_M$ we use their integral
representations in the terms of the world coordinates $z^M$ and $\zeta ^M$
of source particles without AMM.
Starting the construction of the required here action note, that this
functional will also remain invariant for the case when
the superconnection $A_M$ is shifted on the SUSY- and $U(1)$-invariant object
$W'_{M}=(W_{\mu},W'_{\alpha },{\bar W}'{}^{\dot\alpha})$. This object,
in particular, may be constructed of the strength components $F_{MN}$.
As is known \cite{ZheltII}, the inclusion of the interactions of spinning
particles with external electromagnetic field by means of their AMMs
requires using the superstrength components $F_{MN}$. So, it is reasonable
to suppose
that in the considered case an additional shift of the superconnection
\begin{equation}
e A_M \mapsto e A_M +i\mu _1 {W'}_M,
\label{AtoAPlusW}
\end{equation}
may be found sufficient for taking into account the interaction of
superparticles with electromagnetic field via their AMM $\mu $.
The realization of this assumption demands that the components of
$W'_M$ should have proper dimensionalities
$[W'_{\mu}] = L^{-2}$, $[W' _{\alpha }] = [\bar W'{}^{\dot\alpha}] =L^{-3/2}$.
This circumstance, together with the requirement that $W'_M$ must be
presented in the linear form with respect to the superfield
invariant $F_{MN}$, sharply restricts  the possibility to construct
the invariants $W'{}_{M}$.
In particular, the natural Lorentz vector $W' _{\mu}$ with the
dimensionality $L^{-2}$
can not be constructed of $F_{MN}$. At the same time the desired spinor
invariant $W'{}^{\alpha}$ may be taken in the form
$\sim F_{\mu\dot\alpha} \tilde\sigma ^{\mu \dot\alpha \alpha }$.
So, the admissible shift may be chosen as
\begin{equation}
W'_{M} = W_{M} \equiv {i\over 4} (0,-
\sigma _{\mu\alpha{\dot\alpha}}F^{\mu{\dot\alpha}},
{\tilde\sigma}^{\mu{\dot\alpha}\alpha}F_{\mu\alpha}).
\label{WMmy}
\end{equation}
In accordance with this choice, the action (\ref{ActionSUSYIntAndFree}) for
charged superparticles is generalized to the form
\begin{equation}
S_{int}^{(e,\mu )} = i \int
\left[ \omega ^{\mu}(d) e A_{\mu}+
{\theta}^{\alpha}(d)\left( e A_{\alpha} + i\mu _1 W_\alpha\right) +
{\bar\theta}_{\dot\alpha}(d)\left( e {\bar A}^{\dot\alpha} + i\mu _1
{\bar W}^{\dot\alpha}\right)
\right].
\label{SintWithEAndMu}
\end{equation}
In general, $A_M$ and $W_{\alpha} $ in this expression, are the sum of
two integral representations, which describe fields generated by charge
and AMM of particle-source.
It is not so difficult to verify
that this action is symmetric one under permutations
of the particles. Thus it solves the problem of the extension
of the FSTWF approach for the one-half spin
particles with electric charge and AMM.
If we introduce the two-component ``charge''
$q^{\Lambda} \equiv (e,i{\mu })$ and superconnection
$G_M ^{\Lambda} \equiv \left({{\textstyle {A_M}}\atop{\textstyle
{W_M}}}\right)$
this action can be rewritten more compactly
\begin{equation}
S_{int}^{(e,\mu )} = i \int
\omega ^{M}(d) q^{\Lambda} G_{M}^{\Lambda}.
\label{SintWithEAndMuCompact}
\end{equation}
Such an ``isotopic'' form of notation allows to underline
the symmetry between
the pair: $(e, A_M)$ $\leftrightarrow$ $(i\mu , W_M)$.
The considered modification of the superfield connection and the action
leads to the change of the standard supersymmetric and
$U(1)$-covariant derivative $\nabla _M$ into the new $\tilde\nabla _M$
\begin{equation}
\nabla _M \equiv { D}_M + e A_M \ \mapsto \  {\tilde \nabla}_M =
{ D}_M + q^{\Lambda} G_M ^{\Lambda}= {D}_M +eA_M +i\mu  W_M.
\label{MuExtOfDeriv}
\end{equation}
The corresponding change of the standard superfield strengths
$e F_{MN}$  \cite{WESS} into
the extended ones $q^{\Lambda}G^{\Lambda}_{MN}$
is performed following way
\begin{eqnarray}
q^{\Lambda }G_{MN}^{\Lambda}=e F_{MN}+2i\mu D_{[M}W_{N\}}
\label{GForms}
\end{eqnarray}
and permits to rewrite equations of motion of superparticle
with charge and AMM more compactly too.

The physical meaning of the constant $\mu $ as the anomalous magnetic moment
of the particles follows from an analysis of action. After the first pair
of Maxwell equations is taken into account, this action becomes
\begin{eqnarray}
S_{int} ^{(e,\mu )} \Biggr| _{\textstyle e=0, \atop \textstyle{\rm photon}}
&=&i\mu \int d\eta   \left[ (\dot\theta \sigma ^{\mu \nu } \theta )-
(\dot{\bar\theta} \tilde\sigma ^{\mu \nu }\bar\theta )\right] v_{\mu \nu }
\nonumber\\
&&+{1\over 2}\mu \int d\eta   \left[ \theta \theta (\dot \theta \sigma ^{\mu }
\bar\theta ) +\bar\theta \bar\theta (\theta \sigma ^{\mu }\dot{\bar\theta })
\right]\partial ^{\rho }v_{\rho \mu }.
\label{ActMuPhoton}
\end{eqnarray}
Of the two terms remaining in this action, the second describes the
spin-orbit and other relativistic interactions, which correspond to
succeding terms in the expansion in power of $1/c$. Accordingly, we can see
the physical meaning of the constant $\mu $ by restricting the analysis
to the first term. Here it is convenient to switch from the pair of
Weyl spinors $(\theta ,\bar\theta )$ and the matrix $(\sigma ^{\mu \nu })
_{\alpha }^{\beta }$ to the Dirac bispinor $\Psi$ and the spin operator
of the relativistic particle $\Sigma _{\mu \nu }$
\begin{equation}
\Psi =\left( \theta _{\alpha } \atop \dot{\bar\theta }
{}^{\dot\alpha } \right) ,
\qquad \Sigma _{\mu \nu } = {i\over 4} [\gamma _{\mu },\gamma _{\nu } ].
\label{13}
\end{equation}
Here $\gamma _{\mu }$ are the Dirac matrices in Weyl basis. The contribution
of the first term to the action can then be written as a standard Pauli
term
\begin{eqnarray}
S_{int} ^{(e,\mu )} &=&
\mu \int d\tau  \left( \bar\psi  \Sigma ^{\mu \nu } \psi\right)
v_{\mu\nu  } +
\nonumber\\
&&+ \left( \hbox{high-order corrections and other interactions}
\right)
\label{14}
\end{eqnarray}
The physical meaning of the constant $\mu $ is obvious from this
expression: it is the anomalous magnetic moment of the particle,
expressed in Bohr magnetons.

Shown is the principal possibility of the unification of the
action-at-a-distance theory together with the conception of supersymmetry.
This unification permits to generalize the idea of the construction of
fields from world coordinates for spinor fields. As a result, the
fundamental Maxwell and Dirac equations are derived for the fields of the
Maxwell supermultiplet. Another important result of the approach developed
here is the generalization of the minimality principle
taking into account the
electromagnetic interaction via the anomalous magnetic moment of
superparticle. This new form of the minimality principle gives a new
universality relation between the pairs $(e,A_M)$ and $(i\mu ,W_M)$.
The significance of this extended minimality principle falls outside the
frame of the action-at-a-distance theory and may be used as a general
principle in the (super)gauge field theories.


\begin{references}

\bibitem{FOK}
A. D. Fokker, Zeitschrift f{\"u}r Physik {\bf 58}, 386 (1929), and
Phisica {\bf 9}, 83 (1929); Physica {\bf 12}, 145 (1932).

\bibitem{SCH}
K. Schwarzschild, G{\"o}ttinger Nachrichten {\bf 128}, 132 (1903).

\bibitem{TET}
H. Tetrode, Zeitschrift f{\"u}r Physik {\bf 10}, 317 (1922).

\bibitem{FW1}
J. A. Wheeler and R. P. Feynman, Rev. Mod. Phys. {\bf 21}, 425 (1949).


\bibitem{SUPERSTRING}
M. B. Green, J. H. Schwarz and E. Witten,
{\it Superstring theory}
(Cambr. Univ. Press, New York, 1987), Vol. 1 and 2.

\bibitem{Hoyle}
F. Hoyle and J. V. Narlikar, Proc. Roy. Soc. {\bf A277}, No.1 (1964).

\bibitem{Weinberg}
S. Weinberg, {\it Gravitation and Cosmology}
(John Willey \& Jons, NY, 1972)

\bibitem{Kalb}
M. Kalb and P. Ramond,  Phys. Rev. {\bf D9}, 2273 (1974).

\bibitem{WESS}
J. Wess and J. Bagger,  {\it Supersymmetry and supergravity}
(Prin. Univ. Press, Princeton, New Jersey, 1983).

\bibitem{ZT1}
A. A. Zheltukhin, V. V. Tugai,
Supersymmetry and principle of action at a distance,
Pis'ma Zh. Eksp. Teor. Fiz., {\bf 58}, No.4, 246--250 (1993).

\bibitem{ZT3}
V. V. Tugai, A. A. Zheltukhin,
A superfield generalization of the classical action-at-a-distance
theory, Phys. Rev., {\bf D51}, No.8 (1995).

\bibitem{ZheltI}
A. A. Zheltukhin, Phys. Lett. {\bf B168}, No.1,2, 43 (1986);
Yader. Fiz. {\bf 42}, No.3, 720 (1985).

\bibitem{ZheltII}
A. A. Zheltukhin, Teor. Mat. Fiz. {\bf 64}, No.3, 500 (1985).
\vskip 5mm
\end{references}
\end{document}